\documentclass[graybox, envcountchap]{svmult}

\usepackage{graphicx}        % standard LaTeX graphics tool when including figure files

\usepackage[bottom]{footmisc}% places footnotes at page bottom

\usepackage{newtxtext}       % 

\usepackage{url}
\usepackage{booktabs}
\usepackage{hyphenat}

\usepackage{xspace}

\newcommand{\sect}[2]{Section~\ref{#1:sec:#2}}

\newcommand{\fig}[2]{Figure~\ref{#1:fig:#2}}
\newcommand{\tab}[2]{Table~\ref{#1:tab:#2}}

\newcommand*{\eg}{e.g.,\@\xspace}
\newcommand*{\etal}{\emph{et~al.}\@\xspace}

\newcommand{\github}{{GitHub}\xspace}
\newcommand{\gitea}{{Gitea}\xspace}
\newcommand{\gitee}{{Gitee}\xspace}
\newcommand{\gitlab}{GitLab\xspace}
\newcommand{\bitbucket}{BitBucket\xspace}
\newcommand{\sourceforge}{{SourceForge}\xspace}

\newcommand{\docker}{{Docker}\xspace}
\newcommand{\dockerhub}{{Docker Hub}\xspace}

\newcommand{\stackoverflow}{{Stack Overflow}\xspace}

\newcommand{\npm}{{npm}\xspace}

\begin{document}

\title*{An Introduction to Software Ecosystems}
\author{Tom Mens \and Coen De Roover}
\institute{Tom Mens \at University of Mons, Belgium, \email{Tom.Mens@umons.ac.be}
    \and Coen De Roover  \at Vrije Universiteit Brussel, Belgium, \email{Coen.De.Roover@vub.be}}
\maketitle
\label{INT:ch}

%Each chapter should be preceded by an abstract (no more than 200 words) that summarizes the content. The abstract will appear \textit{online} at \url{www.SpringerLink.com} and be available with unrestricted access. This allows unregistered users to read the abstract as a teaser for the complete chapter.
%Please use the 'starred' version of the \texttt{abstract} command for typesetting the text of the online abstracts (cf. source file of this chapter template \texttt{abstract}) and include them with the source files of your manuscript. Use the plain \texttt{abstract} command if the abstract is also to appear in the printed version of the book.}

\abstract*{This chapter defines and presents different kinds of software ecosystems. % that are targeted in this book.
    The focus is on the development, tooling and analytics aspects of ``software ecosystems'', i.e., communities of software developers and the interconnected software components (e.g., projects, libraries, packages, repositories, plug-ins, apps) they are developing and maintaining.
    The technical and social dependencies between these developers and software components form a socio-technical dependency network, and the dynamics of this network change over time.
    We classify and provide several examples of such ecosystems. %, many of which will be explored in further detail in the subsequent chapters of the book.
    The chapter also introduces and clarifies the relevant terms needed to understand and analyse these ecosystems, as well as the techniques and research methods that can be used to analyse different aspects of these ecosystems.}

\abstract{This chapter defines and presents different kinds of software ecosystems. % that are targeted in this book.
    The focus is on the development, tooling and analytics aspects of ``software ecosystems'', i.e., communities of software developers and the interconnected software components (e.g., projects, libraries, packages, repositories, plug-ins, apps) they are developing and maintaining.
    The technical and social dependencies between these developers and software components form a socio-technical dependency network, and the dynamics of this network change over time.
    We classify and provide several examples of such ecosystems. %, many of which will be explored in further detail in the subsequent chapters of the book.
    The chapter also introduces and clarifies the relevant terms needed to understand and analyse these ecosystems, as well as the techniques and research methods that can be used to analyse different aspects of these ecosystems.\\\\    
    {\color{red}{\bf Copyright notice} This is a preprint of the chapter ``An Introduction to Software Ecosystems'' co-authored by Tom Mens and Coen De Roover, published in the book ``Software Ecosystems: Tooling and Analytics'' (eds. Tom Mens, Coen De Roover, Anthony Cleve), 2023, Springer (ISBN 978-3-031-36059-6), reproduced with permission of Springer. The final authenticated version of the book and this chapter is available online at: \url{https://doi.org/10.1007/978-3-031-36060-2}.}}
    
\newpage
%%%%%%%%%%%%%%%%%%%%%%%
%%%%%%%%%%%%%%%%%%%%%%%
\section{The Origins of Software Ecosystems}
\label{INT:sec:origin}
Today, \emph{software ecosystems} are considered an important domain of study within the general discipline of \emph{software engineering}. This section describes its origins, by summarising the important milestones that have led to its emergence.
\fig{INT}{milestones} depicts these milestones chronologically.

\begin{figure}[htbp]
    \centering
    \includegraphics[width=\textwidth]{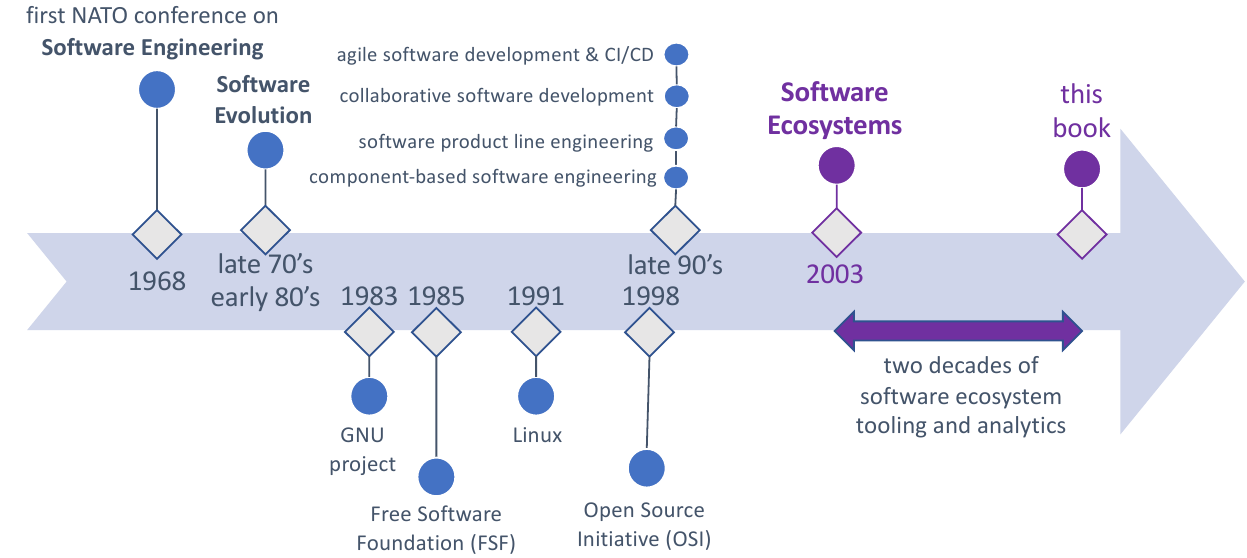} % requires the graphicx package
    \caption{Milestones that contributed to the domain of research (analytics) and development (tooling) of \emph{software ecosystems}}
    \label{INT:fig:milestones}
\end{figure}

The software engineering discipline emerged in 1968 as the result of a first international conference~\cite{Naur1969}, sponsored by the NATO Science Committee, based on the realisation that more disciplined techniques, engineering principles, and theoretical foundations were urgently needed to cope with the increasing complexity, importance, and impact of software systems in all sectors of economy and industry.
Even the key idea of \emph{software reuse}~\cite{Krueger1992,Frakes2005}, which suggests to reduce time-to-market, cost, and effort when building software, while at the same time increasing reuse, productivity, and quality, is as old as the software engineering discipline itself.
During the aforementioned conference, Malcolm Douglas McIllroy proposed to face increasing software complexity by building software through the reuse of high-quality software components~\cite{McIllroy1968}.

In the late seventies, awareness increased that the development of large-scale software needs to \emph{embrace change} as a key aspect of the development process~\cite{Yau1978}.
This has led Manny Lehman to propose the so-called laws of \emph{software evolution}, focusing on how industrial software systems continue to evolve after their first deployment or public release~\cite{Lehman1976,Lehman1980,Lehman1980a}.
The software evolution research domain is still thriving today~\cite{Mens2008,Mens2014-ESS}, with two dedicated annual conferences: the IEEE International Conference on Software Maintenance and Evolution (ICSME) and the IEEE Software Analysis, Evolution and Reengineering Conference (SANER).

Another important factor having contributed to the popularity of software ecosystems is the emergence and ever-increasing importance of \emph{free software} and \emph{open source software} (OSS) since the early eighties, partly through the creation of the GNU project\footnote{\url{https://www.gnu.org}} in 1983 and the Free Software Foundation (FSF) in 1985 by Richard Stallman, as well as the creation of the Linux operating system in 1991.
Strong open source advocates such as Eric Raymond~\cite{Raymond1999} further contributed to the popularity through the creation of the Open Source Initiative (OSI) in 1998, and by contrasting cathedral-style closed development process models with the bazaar-style open development process models for open source and free software in which the code is publicly developed over the Internet.
This bazaar-style model evolved into geographically distributed \emph{global software development}~\cite{Grinter1999,Herbsleb2001} models, supported by the immensely popular \emph{social coding platforms}~\cite{Dabbish2012} such as GitHub, GitLab, Gitea and BitBucket.

In parallel, the importance of software reuse in the late nineties gave rise to additional subfields of software engineering such as the domain of \emph{component-based software engineering}~\cite{Szyperski1997,Kozaczynski1998CBSE}, focusing on methods and principles for composing large systems from loosely-coupled and independently-evolving software components.
Around the same time it was joined by another subfield, called \emph{software product line engineering}~\cite{Weiss199SPLE,Clements1999}, which explicitly aims to enable developing closely-related software products using a process modelled after product line manufacturing, separating the \emph{domain engineering} phase of producing reusable software artefacts that are common the product family, from the \emph{application engineering} phase that focuses on developing concrete software applications that exploit the commonalities of the reusable artefacts created during the domain engineering phase.
Software product lines have allowed many companies to reduce costs while at the same time increasing quality and time to market, by providing a product line platform and architecture that allows to scale up from the development and maintenance of individual software products to the maintenance of entire families of software products. However, these product families still remain within the organisational boundaries of the company.

Around the same time, the lightweight and iterative process models known as \emph{agile software processes} started to come to the forefront, with a user-centric vision requiring adaptive and continuous software change.
Different variants, such as Scrum \cite{Schwaber1997} and eXtreme Programming (XP) \cite{Beck1999}, led to the foundation of the Agile Alliance and the creation of the agile manifesto \cite{beck2001manifesto}.
In support of agile software processes, various development practices and tools for \emph{continuous integration and delivery} (CI/CD) emerged later on in the decade.

Since the seminal 2003 book by Messerschmitt and Szyperski~\cite{messerschmitt2003software}, \emph{software ecosystems} have become an active topic of research in software engineering.
As argued by Jan Bosch~\cite{Bosch2009,Bosch2010}, software ecosystems expand upon software product lines by allowing companies to cross the organisational boundaries and make their software development platforms available to third parties that, in turn, can contribute to the popularity of the produced software through externally developed components and applications.
The key point of software ecosystems is that software products can no longer be considered or maintained in isolation, since they have become heavily interconnected.

\section{Perspectives and Definitions of Software Ecosystems}
\label{INT:sec:definition}

Messerschmitt and Szyperski \cite{messerschmitt2003software} were arguably among the first to use the term \emph{software ecosystem}, and defined it rather generically as \emph{``a collection of software products that have some given degree of symbiotic relationships.''} Since then, the research literature has provided different definitions of software ecosystems, from many different perspectives.

From an \textbf{ecological perspective}, several researchers have tried to exploit the analogy between software ecosystems and natural ecosystems.
The term software ecosystem quite obviously originates from its ecological counterpart of biological ecosystems that can be found in nature, in a wide variety of forms (\eg rainforests, coral reefs, deserts, mountain zones, and polar ecosystems).
In 1930, Roy Clapham introduced the term \emph{ecosystem} in an ecological context to denote the \emph{``physical and biological components of an environment considered in relation to each other as a unit''}~\cite{Willis1997}.
These components encompass all living organisms (e.g., plants, animals, micro-organisms) and physical constituents (e.g., light, water, soil, rocks, minerals) that interact with one another in a given environment.
Dunghana \etal \cite{Dhungana2013} compared the characteristics of natural and software ecosystems. Mens \cite{Mens2015} provided a high-level historical and ecological perspective on how software ecosystems evolve.
Moore \cite{Moore1993} and Iansiti and Levien \cite{Iansiti2004} focused on the analogy between business ecosystems and ecology.

From an \textbf{economic and business perspective}, Jansen \etal \cite{Jansen2009-ICSE} provide a more precise definition: \emph{``a set of businesses functioning as a unit and interacting with a shared market for software and services, together with the relationships among them.''} In a similar vein,
Bosch \etal~\cite{Bosch2009} say that a software ecosystem \emph{``consists of a software platform, a set of internal and external developers and a community of domain experts in service to a community of users that compose relevant solution elements to satisfy their needs.''}
Hanssen~\cite{Hanssen2012} defines it as \emph{``a networked community of organizations, which base their relations to each other on a common interest in a central software technology.''}
An excellent entry point to this business-oriented viewpoint on software ecosystems is the book edited by Jansen \etal~\cite{Jansen2013book}.
%In contrast, the chapters in the current book focus mostly on the complementary technical and social perspectives.

From a more \textbf{technical perspective}, the focus is on technical aspects such as the software tools that are being used (\eg version control systems, issue and bug trackers, social coding platforms, integrated development environments, programming languages) and the software artefacts that are being used and produced (\eg source code, executable code, tests, databases, documentation, trace logs, bug and vulnerability reports).
Within this technical perspective, Lungu \cite{Lungu2008} defined a software ecosystem as \emph{``a collection of software projects that are developed and evolve together in the same environment''.}
The notion of \emph{environment} can be interpreted rather broadly.
The environment can correspond to a software-producing organisation, including the tools and libraries used by this organisation for developing its software projects, as well as the clients using the developed software projects.
It can correspond to an academic environment, composed of software projects developed and maintained by students and researchers in research units.
It can also correspond to an entire OSS community consisting of geographically dispersed project collaborators focused around similar philosophies or goals.

From a \textbf{social perspective}, the focus is on the social context and network structure that emerges as a result of the collaboration dynamics and interaction between the different contributors to the projects that belong to the software ecosystem.
This social structure is at least as important as the technical aspects, and includes the various stakeholders that participate in the software ecosystem, such as developers, end users, project managers, analysts, designers, software architects, security specialists, legal consultants, clients, QA teams, and many more.
%\chap{EMO} focuses on these social aspects from an emotion analysis viewpoint.

Manikas \cite{ManikasHansen2013} combined all these perspectives into a single all-encompassing definition of a software ecosystem as \emph{``the interactions of a set of actors on top of a common technological platform that results in a number of software solutions or services. Each actor is motivated by a set of interests or business models and connected to the rest of the actors and the ecosystem as a whole with symbiotic relationships, while the technological platform is structured in a way that allows the involvement and contribution of the different actors.''}

%(Terminology of "registry" / "index")
%(Notion of reusable assets / components / dependencies )

\section{Examples of Software Ecosystems}
\label{INT:sec:SECO-examples}

Following the wide diversity of definitions of software ecosystem, the kinds of software ecosystems that have been studied in recent research are equally diverse.
An interesting entry point into how the research literature on software ecosystems has been evolving over the years are the many published systematic literature reviews, such as \cite{Barbosa2011, ManikasHansen2013, Manikas2016, Seppanen2017, Burstrom2022}.

Without attempting to be complete, \tab{INT}{secokinds} groups into different categories some of the most popular examples of software ecosystems that have been studied in the research literature. These categories are not necessarily disjoint, since software ecosystems tend to contain different types of components that can be studied from different viewpoints.

% Requires the booktabs if the memoir class is not being used
\begin{table}[htbp]
    \centering
    %\topcaption{Table captions are better up top} % requires the topcapt package
    \caption{Categories of software ecosystems}
    \label{INT:tab:secokinds}
    \begin{tabular}{p{2.2cm}p{3.3cm}p{2.8cm}p{2.7cm}}
        \toprule
        Category                           & Examples                                                                                                                                         & Components                                                                               & Contributors                                                                                  \\
        \midrule
        \begin{flushleft}\nohyphens{digital platforms}\end{flushleft}                  & \begin{flushleft}\nohyphens{mobile app stores, integrated development environments}\end{flushleft}                                                                                          & \begin{flushleft}mobile apps, software plug-ins or extensions\end{flushleft}                                             & \begin{flushleft}third-party app or plug-in developers and their users\end{flushleft}                                         \\
        \begin{flushleft}\nohyphens{social coding platforms}\end{flushleft}             & \begin{flushleft}\nohyphens{\sourceforge, \github, \gitlab, \gitea, \bitbucket}\end{flushleft}                                                                                               & \begin{flushleft}\nohyphens{software project repositories}\end{flushleft}                                                            & \begin{flushleft}\nohyphens{software project contributors}\end{flushleft}                                                                 \\
        \begin{flushleft}\nohyphens{component-based software ecosystems}\end{flushleft} & \begin{flushleft}\nohyphens{software library registries (\eg CRAN, \npm, RubyGems, PyPi, Maven Central), OS package registries (\eg Debian packages, Ubuntu package archive)}\end{flushleft} & \begin{flushleft}\nohyphens{interdependent software packages}\end{flushleft}                                                         & \begin{flushleft}\nohyphens{consumers and producers of software packages and libraries}\end{flushleft}                                    \\
        \begin{flushleft}\nohyphens{software automation ecosystems}\end{flushleft}     & \begin{flushleft}\nohyphens{\dockerhub, Kubernetes, Ansible~Galaxy, Chef~Supermarket, Puppetforge}\end{flushleft}                                                                            & \begin{flushleft}\nohyphens{container images, configuration and orchestration scripts, CI/CD pipelines and workflows}\end{flushleft} & \begin{flushleft}\nohyphens{creators and maintainers of workflow automation, containerisation and orchestration solutions}\end{flushleft} \\
        \begin{flushleft}\nohyphens{communication-oriented ecosystems}\end{flushleft}   & \begin{flushleft}\nohyphens{mailing lists, Stack~Overflow, Slack}\end{flushleft}                                                                                                             & \begin{flushleft}\nohyphens{e-mail threads, questions, answers, messages, posts, \ldots}\end{flushleft}                              & \begin{flushleft}\nohyphens{programmers, developers, end-users, researchers}\end{flushleft}                                               \\
        \begin{flushleft}\nohyphens{OSS communities}\end{flushleft}                  & \begin{flushleft}\nohyphens{Apache Software Foundation, Linux Foundation}\end{flushleft}                                                                                                     & \begin{flushleft}OSS projects\end{flushleft}                                                                            &  \begin{flushleft}\nohyphens{community members, code contributors, project maintainers, end users}\end{flushleft}                             \\
        \bottomrule
    \end{tabular}
\end{table}

The remaining subsections provide more details for each category, illustrating the variety of software ecosystems that have been studied, and providing examples of well-known ecosystems and empirical research that has been conducted on them.

\subsection{Digital Platform Ecosystems}
\label{INT:sec:digitalplatforms}

Hein \etal~\cite{Hein2020} define a \emph{digital platform ecosystem} as a software  ecosystem that \emph{``comprises a platform owner that implements governance mechanisms to facilitate value-creating mechanisms on a digital platform between the platform owner and an ecosystem of autonomous complementors and consumers''}.
This is in line with the previously mentioned definition by Bosch \etal~\cite{Bosch2009} that a software ecosystem \emph{``consists of a software platform, a set of internal and external developers and a community of domain experts in service to a community of users that compose relevant solution elements to satisfy their needs.''}

Well-known examples of digital platform ecosystems are the \emph{mobile software ecosystems} provided by companies such as Microsoft, Apple and Google.
The company owns and controls an \emph{app store} as a central platform to which other companies or individuals can contribute apps, which in turn can be downloaded and installed by mobile device users.
The systematic mapping studies by de Lima Fontao \etal \cite{deLimaFontao2015} and \cite{Steglich2019} report on the abundant research that has been conducted on these mobile software ecosystems.

Any software system that provides a mechanism for third parties to contribute plug-ins or extensions that enhance the functionalities of the system can be considered as a digital software ecosystem. Examples of these are configurable text editors such as Emacs and Vim, and integrated software development environments (IDEs) such as IntelliJ IDEA, VS Code, NetBeans and Eclipse.
The latter ecosystem in particular has been the subject of quite some research on its evolutionary dynamics (\eg \cite{MensRamil2008-ICSM,Businge:SQJ:2015,Businge2012Survival,Businge2013CSMR,Businge:Eclise:saner:2019,Kawuma:ICPC:2016,2-236, AbouKhalil2021, Nugroho2021}).
These examples show that digital platform ecosystems are not necessarily controlled by a single company.
In many cases, they are managed by a consortium, foundation or open source community.
For example, NetBeans is controlled by the Apache Foundation, and Eclipse is controlled by the Eclipse Foundation.

Another well-known digital platform ecosystem is WordPress, the most popular content management system in use today, which features a plugin architecture and template system that enables third parties to publish themes and extend the core functionality.
Um \etal \cite{Um2022WordPress} presented a recent study of this ecosystem.
Yet another example is OpenStack, an open source cloud computing platform involving more than 500 companies.
This ecosystem has been studied by several researchers (\eg \cite{2-236,Teixeira2017,Foundjem:2022wx,Zhang2022}).

\subsection{Component-based Software Ecosystems}
\label{INT:sec:CBSECO}

A very important category of software ecosystems are so-called \emph{component-based software ecosystems}.
They constitute large collections of reusable software components, which often have many interdependencies among them~\cite{Abate2009}.
Empirical studies on component-based software ecosystems tend to focus on the technicalities of dependency-based reuse, which differentiates them from studies on digital platform ecosystems which have a more business-oriented and managerial focus.

As explained in \sect{INT}{origin}, the idea of building software by reusing existing software components is as old as the software engineering discipline itself, since it was proposed by McIllroy in 1968 during the very first software engineering conference~\cite{McIllroy1968}.
The goal was to reduce time-to-market, cost and effort when building software, while at the same time increasing reuse, productivity and quality.
This has given rise to a very important and abundant subfield of software engineering that is commonly referred to as component-based software engineering.
Despite the large body of research in this field (\eg \cite{Caldiera1991,Krueger1992,Szyperski1997}) it was not able to live up to its promises due to a lack of a standard marketplace for software components, combined with a lack of proper component models, terminology, and scalable tooling \cite{Kotovs2009}.
All of this has changed nowadays, probably due to a combination of the increasing popularity of OSS and the emergence of affordable cloud computing solutions.

Among the most important success stories of component-based software ecosystems are undoubtedly the
many interconnected \emph{software packages} for OSS operating systems such as the GNU Project since 1983, Linux since 1991, Debian since 1993 (\eg \cite{Abate2009,gonzalez2009macro,debsources-esem-2014,Claes2015,Claes2018}) and Ubuntu since 2004.
They come with associated \emph{package management systems} (or \emph{package managers} for short) such as DPKG (since 1994) and APT (since 1998), which are systems that automate the process of selecting, installing (or removing), upgrading and configuring of those packages.
Package managers typically maintain a database of software dependencies and version information to prevent software incompatibilities.

Another popular type of ecosystems of reusable components are \emph{software libraries}. Software developers, regardless of whether they are part of an OSS community or software company, rely to a large extent on such reusable third-party \emph{software libraries}. These library ecosystems tend to come with their own specific package managers and package registries, and are available for all major programming languages. Examples include the CPAN archive network (created in 1995 for the Perl programming language, the CRAN archive network (created in 1997) and Bioconductor for the R statistical programming language)~\cite{German2013,Plakidas2017-JSS},
npm %(since 2010)
and Bower for JavaScript~\cite{cogo2021deprecation, abdalkareem2017developers,decan2018evolution,decan2018impact,decan2021back,zerouali2022impact}, %ahmed citeren, kth citeren
PyPI for Python~\cite{valiev2018ecosystem}, %(since 2003)
Maven (Central) for JVM-based languages such as Java and Scala~\cite{Benelallam2019,soto2021comprehensive,Ochoa2022-EMSE},
Packagist for PHP,
RubyGems for Ruby~\cite{Kabbedijk2011,decan2021back,zerouali2022impact},
NuGet for the .NET ecosystem~\cite{Li2022-ICSE},
and the Cargo package manager and its associated crates registry for the Rust programming language~\cite{decan:emse:2019,Schueller2022-Rust}.
Another example is the Robot Operating System (ROS), the most popular middleware for robotics development, offering reusable libraries for building a robot, distributed through a dedicated package manager \cite{Estefo2019,Pichler2019ROS,Kolak2020ROS}.

Decan \etal \cite{decan:emse:2019} studied and compared seven software library ecosystems for programming languages, focusing on the evolutionary characteristics of their package dependency networks.
They observed that library dependency networks tend to grow over time, but that some packages are more impactful than other.
A minority of packages are responsible for most of the package updates, a small proportion of packages accounts for most of the reverse dependencies, and there is a high proportion of fragile packages due to a high number of transitive dependencies.
This makes software library ecosystems prone to a variety of technical, dependency-related issues~\cite{decan2017empirical,abdalkareem2017developers,Claes2018,soto2021comprehensive}, licensing issues~\cite{2022:icsr:makari}, security vulnerabilities~\cite{decan2018impact,zerouali2022impact,Alfadel2021}, backward compatibility~\cite{decan2021what,decan2021back,bogart2021and}, and reliance on deprecated components~\cite{cogo2021deprecation}, as well as obsolete or outdated components \cite{decan2018evolution,zerouali2019formal,lauinger2018thou,Stringer2020}.
Versioning practices, such as the use of semantic versioning, can be used to a certain extent to reduce some of these risks~\cite{dietrich2019dependency,lam2020putting,decan2021what,Ochoa2022-EMSE}.
Library ecosystems also face many social challenges, such as how to attract and retain contributors and how to avoid contributor abandonment \cite{Constantinou2017}.

%Smalltalk/Pharo: robbes2012developers, horadevelopersapievolution

\subsection{Web-based Code Hosting Platforms}

\begin{figure}
    \centering
    \includegraphics[width=\textwidth]{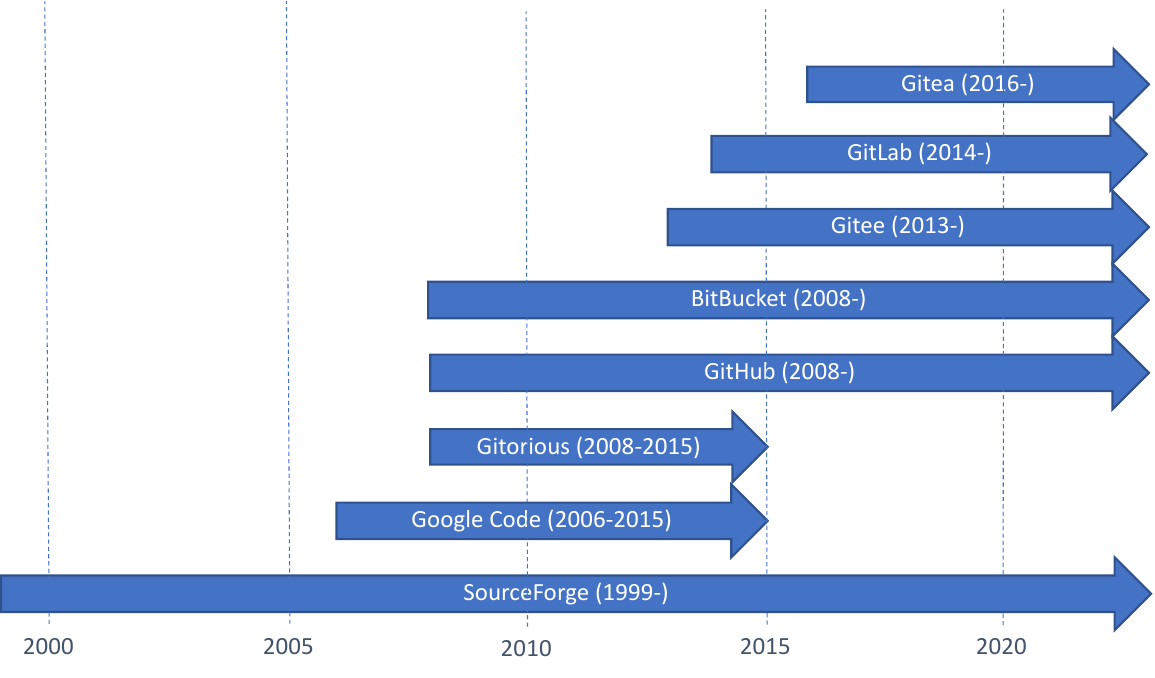} % requires the graphicx package
    \caption{Historical overview of source code hosting platforms}
    \label{INT:fig:codehosting}
\end{figure}

The landscape of web-based code hosting platforms has seen many important changes over the last two decades, as can be seen in \fig{INT}{codehosting}.
\sourceforge was created in 1999 as a centralized web-based platform for hosting and managing the version history of free and OSS projects.
It used to be a very popular data source for empirical research (\eg \cite{Howison2004,Robles2006, Koch2009,Nyman2011}).
This is no longer the case today, as the majority of OSS projects have migrated to other hosting platforms.
Google started running a similar open source project hosting service, called Google Code, in 2006 but shut it down in January 2016.
The same happened to Gitorious which ran from 2008 to 2015.

\github replaced Google Code as the most popular and largest hosting platform for open source (and commercial) software projects that use the git version control system.
Other alternatives such as \bitbucket (also created in 2008) and \gitlab (created in 2014) and the likes are much less popular for hosting OSS projects.
Even older is \gitee (created in 2013), an online git forge mainly used in China for hosting open source software.
A relatively new contender in the field is \gitea, created in 2016 and funded by the Open Source Collective.
%For completeness we also mention Software Heritage that will be presented in detail in \chap{SWH}. Strictly speaking it is not a code hosting platform, but rather a code archival platform aiming to preserve software code in the very long term.

\github maintains historical information about hundreds of millions of OSS repositories and has been the subject of many empirical studies focusing on different aspects \cite{Kalliamvakou2014}.
\github is claimed to be the first \emph{social coding} platform~\cite{Dabbish2012}, as it was the first hosting platform to provide a wide range of mechanisms and associated visualisations to increase collaboration by making socially significant information visible: watching, starring, commits, issues, pull requests and commenting.
Being an enabler of social coding, the social aspects around \github projects have been studied extensively~\cite{Tsay2014, padhye2014extcontrib}, including
communication patterns \cite{1-042},
collaboration through pull requests \cite{Rahman:MSR:2014, Yu:MSR:2015, Gousios2016},
variation in contributor workload~\cite{Vasilescu2014},
gender and tenure diversity \cite{vasilescu2015gender,vasilescu2015quality},
geographically distributed development \cite{takhteyev2010ossgeography, rastogi2018geobias, wachs2021ossgeography},
socio-technical alignment between projects~\cite{Blincoe2019},
the impact of gamification on the behaviour of software developers~\cite{Moldon2021},
and sentiment and emotion analysis \cite{1-011, 1-063, 1-067, 1-055, 1-076, Guzman:2014:SAC:2597073.2597118}.
%The latter will be presented in more detail in \chap{EMO}.
%
The phenomenon of project forking has also been actively studied in the context of \github  \cite{biazzini2014maythefork, Jiang:emse:2017, Zhou:ICSE:2020}.%, as will be discussed in more detail in \chap{FRK}.
The automation of development activities in \github projects has also been studied, such as the use of CI/CD tools \cite{vasilescu2015quality, beller2017oops, Golzadeh2021SANER}, and the use of development bots \cite{Bodegha2021,Wang2022-butler, AbdellatifBotHunter2022, wessel2022emse}.
%The latter perspective on the ecosystem will be discussed in \chap{WFA}.
The same chapter also explains how \github can be studied from the point of view of a digital platform ecosystem (cf. \sect{INT}{digitalplatforms}), as it offers a MarketPlace of Apps and Actions that can be provided by third-parties.

%TOM: References not included in the above:
%\begin{itemize}
%\item code quality: \cite{ray2017codequality}
%\item README files: \cite{prana2019readme}
%\end{itemize}

\subsection{Open Source Software Communities}

Quite some research on software ecosystems has focused on collections of OSS projects maintained by decentralised communities of software developers.
Such OSS ecosystems have clear advantages over closed, proprietary software ecosystems.
For example, their openness guarantees the accessibility to all.
Following the adagio that ``given enough eyeballs, all bugs are shallow'' \cite{Raymond:Cathedral:2001}, OSS ecosystems benefit from a potentially very large number of people that can report bugs, review the code and identify potential security issues.
Provided that the software licences being used are compatible, organisations and companies can save money by relying on OSS components rather than reinventing the wheel and developing those components themselves.

At the downside, OSS ecosystems and their constituent components are frequently maintained on a volunteer basis by unpaid developers.
This imposes an increased risk of unmaintained components or slow response time.
Organisations that rely on OSS ecosystems could significantly reduce these risks by financially sponsoring the respective communities of OSS developers. Many fiscal and legal initiatives for doing so exist, such as the Open Collective, the Open Source Collective, and  the Open Collective Foundation.

OSS ecosystems are often controlled, maintained and hosted by a non-profit software foundation.
A well-known example is the \emph{Apache Software Foundation} (\url{www.apache.org}).
It hosts several hundreds of OSS projects, involving tens of thousands of code contributors.
This ecosystem has been a popular subject of research (\eg \cite{Bavota2013-ICSM, Chen:2017:CLP:3042021.3042046, Tan2020, Mockus:TOSEM:2002, Calefato2019-IST}).
Another example is the \emph{Linux Foundation} (\url{www.linuxfoundation.org}), whose initial goal was to support the development and evolution of the Linux operating system, but nowadays hosts hundreds of OSS projects with hundreds of thousands of code contributors.
As can be expected, the OSS project communities of third-party components that surround a digital platform ecosystem (cf. \sect{INT}{digitalplatforms}) also tend to be managed by non-profit foundations.
For example, the Eclipse Foundation controls the Eclipse plug-ins, the WordPress Foundation controls the WordPress plug-ins, and the Open Infrastructure Foundation manages the OpenStack projects.

Much in the same way as public OSS ecosystems, there exists a multitude of entirely private and company-controlled software ecosystems. We defer to the book by Jansen \etal~\cite{Jansen2013book} that focuses on the business aspects of such commercial software ecosystems.
Given their proprietary nature they have been much less the subject of quantitative empirical research studies, but it is likely that such private ecosystems share many of the characteristics known to OSS ecosystems. As a matter of illustration of such ecosystems, among the countless available examples we mention Wolfram's Mathematica and MathWorks' MATLAB with their large collections of --often third-party-- add-ons, and the ecosystem surrounding SAP, the world's largest software vendor of enterprise resource planning solutions.

\subsection{Communication-Oriented Ecosystems}

The previous categories of software ecosystems have in common that the main components they focus on are \emph{technical} code-related software artefacts (\eg software library packages and their metadata, mobile software applications, software plug-ins, project repositories, application code and tests, software containers, configuration scripts).

The current category focuses on what we will refer to as \emph{communication-oriented ecosystems}, in which the main component is some \emph{social} communication artefact that is shared among members of a software community through some communication channel.
Examples of these are mailing lists, developer discussion fora, Question and Answer (Q\&A) platforms such as Stack Overflow, and modern communication platforms such as Slack and Discord. %and community-maintained wiki pages such as Wikipedia.
Each of them constitute software ecosystems in their own right.
A particularity of these ecosystems is that the main components they contain (\eg questions, answers, posts, e-mail and message threads) are mostly based on unstructured or semi-structured text.
As a consequence, extracting and analysing relevant information from them requires specific techniques based on Natural Language Processing (NLP).
These ``social programmer ecosystems'' \cite{NovielliCL15} have been analysed by researchers for various reasons, mostly from a social viewpoint:

\smallskip
\textbf{Mailing lists.} Mailing lists are a common communication medium for software development teams, although they are gradually being replaced by more modern communication technologies.
As the same person may have multiple email addresses, disambiguation techniques are often required to uniquely identify a given team member \cite{wiese2016mailing}.
They have been the subject of multiple empirical studies (\eg
\cite{Guzzi2013-MSR,Zagalsky2018-EMSE}).
Some of these studies have tried to identify personality traits or emotions expressed through e-mails \cite{1-082, 2-172, RigbyHassan}.

\smallskip
\textbf{Discussion fora.} Software development discussion fora support mass communication and coordination among distributed software development teams~\cite{Storey2017-TSE}.
They are a considerable improvement over mailing lists in that they provide browse and search functions, as well as a platform for posting questions within a specific domain of interest and for receiving expert answers to these questions.\\
A generic, and undoubtedly the most popular, discussion forum is Stack Overflow, dedicated to questions and answers related to computer programming and software development.
It belongs to the Stack Exchange network, providing a range of websites covering specific topics.
Such Q\&A platforms can be considered as a software ecosystem where the ``components'' are questions and their answers (including all the metadata that comes with them), and the contributor community consists of developers that are asking questions, and experts that provide answers to these questions.
The StackOverflow ecosystem has been studied for various purposes and in various ways~\cite{SO1,SO3,SO2,Nagy2015,Zagalsky2018-EMSE,Bangash2019,Manes2019,AhasanuzzamanAR20,2021:msr:zerouali,2022:saner:velazquez,2023:scico:velazquez}.
The open dataset SOTorrent has been made available on top of a datadump with all posts from 2018 till 2020 \cite{Baltes2018,Baltes2019,baltes2021zenodo}.
Some researchers \cite{NovielliCL15,FontaoESD17,2-083} have applied sentiment and emotion analysis techniques on data extracted from \stackoverflow.
%We refer to \chap{EMO} for a more detailed account on the use of such techniques in software ecosystems.

Next to generic discussion fora such as Stack Overflow, some software project communities prefer to use project-specific discussion fora.
This is for example the case for Eclipse.
Nugroho \etal~\cite{Nugroho2021} present an empirical analysis of how this forum is being used by its participants.

\smallskip
\textbf{Modern communication platforms.} Several kinds of modern communication platforms, such as Slack and Discord are increasingly used by software development teams.
Lin \etal~\cite{Lin2016} reported how Slack facilitates messaging and archiving, as well as the creation of automated integrations with external services and bots to support the work of software development teams.

%\smallskip
%TOM: There is not much to say about Wikipedia, and currently there is only one not very convincing reference, that is more about schema evolution, so it is not relevant in the context of this section. Therefore, it is probably better not to mention Wikipedia here...
%\textbf{Wikipedia} \cite{Curino2008} \tom{To be extended}

\subsection{Software Automation Ecosystems}

Another category of software ecosystems is what we would refer to as software automation ecosystems.
They revolve around technological solutions that aim to automate part of the management, development, packaging, deployment, delivery, configuration and orchestration of software applications, often through cloud-based platforms. We can mention at least three categories: containerisation solutions, orchestration tools based on Infrastructure as Code, and  tools for automating DevOps and CI/CD.

\textbf{Containerisation.} Containerisation allows developers to package all (software and data) components of their applications into so-called containers, which are lightweight, portable and self-contained
executable software environments that can run on any operating system or cloud platform.
By isolating the software applications from the underlying hardware infrastructure, they become easier to manage and more resilient to change.
\docker is the most popular containerisation tool, and it comes with multiple online registries to store, manage, distribute and share containers (\eg Google Container Registry, Amazon ECR, JFrog Container Registry, RedHat's Quay, and of course \dockerhub).
While each of these registries come with their own set of features and benefits, \dockerhub is by far the largest of these registries.
%The corresponding ecosystem is studied in \chap{IAC}, and more specifically in \sect{IAC}{docker}.

\textbf{Management.}
Through \emph{Infrastructure as Code} (IaC), infrastructure management tools enable automating the provisioning, configuration, deployment, scaling and load balancing of the machines used in a digital infrastructure.
Different infrastructure management tools have been proposed, including Ansible, Chef and Puppet.
Each of them come with their own platform or registry for sharing configuration scripts (Ansible Galaxy, Chef Supermarket and PuppetForge).
Sharma \etal\cite{2016:msr:sharma} studied best practices in Puppet configuration code, analysing 4,621 Puppet repositories for the presence of implementation and design configuration smells.
Opdebeeck \etal conversely studied variable-related~\cite{2022:msr:opdebeeck} and security-related~\cite{2023:msr:opdebeeck} bad smells in Ansible files respectively.
The Ansible Galaxy ecosystem has been an active subject of study in general.%, as will be shown in \chap{IAC}, and more specifically \sect{IAC}{ansible}.

\textbf{DevOps and CI/CD.} Collaborative distributed software development processes, especially for software projects hosted on social coding platforms, tend to be streamlined and automated using continuous integration, deployment and delivery tools (CI/CD), which are a key part of DevOps practices.
CI/CD tools enable project maintainers to specify project-specific workflows or pipelines that automate many repetitive and error-prone human activities that are part of the development process.
Examples are test automation, code quality analysis, dependency management, and vulnerability detection.
A wide range of CI/CD tools exist (\eg Jenkins, Travis, CircleCI, GitLab CI/CD and GitHub Actions to name just a few). Coming with a registry or marketplace of reusable workflow components that facilitate the creation and evolution of workflows, ecosystems have formed around many of these tools, such as the ecosystem of GitHub Actions, the integrated CI/CD service of \github.
Since its introduction, the CI/CD landscape on GitHub has radically changed \cite{kinsman2021software, decanuse, wessel2022github}.

%%%%%%%%%%%%%%%%%%%%%%%
%%%%%%%%%%%%%%%%%%%%%%%
\section{Data Sources for Mining Software Ecosystems}

The Mining Software Repositories (MSR) research community relies on a wide variety of publicly accessible raw data, APIs or other data extraction tools, data dumps, curated datasets, and data processing tools (\eg dedicated parsers) depending on the specific purpose and needs of the research being conducted.

\smallskip \textbf{The pros.}
These data sources and their associated tooling form a gold mine for empirical research in software engineering, and they have allowed the MSR field to thrive.
Relying on existing publicly accessible data substantially reduces the laborious and error-prone effort of the data extraction and processing phases of empirical research.
As such, it has allowed researchers and software practitioners to learn a great deal about software engineering practices in the field, and how to improve these practices.
Moreover, this allows multiple researchers to rely on the same data, facilitating comparison and reproducibility of research results~\cite{Gonzalez-Barahona2012}.

\smallskip \textbf{The cons.}
At the same time, these data sources and tools come with a variety of negative consequences, such as:
\begin{itemize}
    \item Existing data and tools can quickly become obsolete, as it is difficult and effort-intensive to keep up with changes in the original data source or in the APIs required to access them. Many initiatives to create and maintain data extraction tools or curated datasets have been discontinued, mainly due to a lack of continued  funding or because the original maintainers have abandoned the initiative due to career changes.

    \item Ethical, legal, or privacy reasons may prevent specific parts of the data of interest to be made available~\cite{gold2022ethicsmsr}. Examples are proprietary copyrighted source code, or personal information that cannot be revealed due by GDPR regulations.

    \item Specific analyses may need specific types of data that are not readily available in existing datasets,  requiring the creation of new datasets or the extension of existing ones. Talking from a personal experience, it often takes several months of effort to obtain, preprocess, validate, curate and improve the quality of the obtained data. Not doing so may lead to results that are inaccurate, biased, not replicable, or not generalisable to other situations.

    \item Existing datasets may not be appropriate for specific analyses, because of how the data has been gathered or filtered.
          As an illustration of this problem, suppose for example that we want to analyse the effort spent by human contributors in some software ecosystem, based on an available dataset containing contributor accounts and their associated activities over time. If this dataset does not distinguish between human accounts and automated bots, then the results will be biased by bot activities being considered as human activities, calling for the use of bot identification approaches and associated datasets (\eg \cite{Bodegha2021}).

    \item Research that is relying on raw data sources instead of curated datasets may reduce reproducibility since, unlike for a published dataset, there is no guarantee that the original data will remain the same after publication of the research results. For example, \github repositories may be deleted and the history of a git repository may be changed at any time~\cite{Bird2009,Kalliamvakou2016}.
\end{itemize}

%\tom{List of datasets that have been made available for use in past and current research on analysing software  ecosystems. There are many such datasets (most being obsolete by now), so it is virtually impossible to be complete.}
%\tom{@Coen: suggest to summarise these datasets in the form of a table? (Columns: name of dataset; citation of dataset; contents of dataset; period/duration of data dumps) Perhaps this is not possible since too diverse and too many datasets.}

The following subsections provide a list of data sources that have been used in empirical research on a wide variety of software ecosystems.
This list is non-exhaustive, given the plethora of established and newly-emerging ecosystems, data sources about them, and research studies on them.

\subsection{Mining the \github Ecosystem}

For git repositories hosted on the \github social coding platform, different ways have been proposed to source their data.
\github provides public REST and GraphQL APIs to interact with its huge dataset of events and interaction with the hosted repositories.
As an alternative, different attempts have been made to provide datasets and data dumps containing relevant data extracted from \github, with varying success:

\begin{itemize}

    \item \emph{GHArchive}\footnote{\url{https://www.gharchive.org}} records, archives and makes available the public \github timeline for public consumption and analysis. It is available on Google BigQuery and it contains datasets, aggregated into hourly archives, based on 20+ event types, ranging from new commits and fork events, to opening new tickets, commenting, and adding members to a project.

    \item In a similar way, \emph{GHTorrent} aimed to obtain data from \github public repositories \cite{GHTorrent,Gousios2017}, covering a large part of the activity from 2012 till 2019.
          The latest available data dump was created in March 2021,\footnote{\url{http://ghtorrent-downloads.ewi.tudelft.nl/mysql/}} and the initiative has been discontinued altogether.

    \item \emph{TravisTorrent} was a dataset created in 2017 based on Travis CI and GitHub, It provides access to over 2.6 million Travis builds from more than 1,000 GitHub projects \cite{Beller2017-MSR}.

\end{itemize}

\subsection{Mining the Java Ecosystem}

Multiple datasets have been produced for use in studies on the ecosystem surrounding the Java programming language.
The Qualitas Corpus~\cite{Tempero2010QualitasCorpus}, a curated dataset of Java software systems, aimed to facilitate reproducing these studies.
Only two datadumps have been released, in 2010 and in 2013.

More recent datasets for Java focused on Apache's Maven Central Repository, a software package registry maintaining a huge collection of libraries for the Java Virtual Machine.
For example, Raemaekers \etal provide the Maven Dependency Dataset with metrics, changes, and a dependency graph for 148,253 jar files \cite{Raemaekers2013}.
The dataset was used to study the phenomena of semantic versioning and breaking changes~\cite{raemaekers2014semantic}.
Mitropoulos \etal \cite{Mitropoulos2014} provide a complementary dataset containing the FindBugs results for every project version included in the Maven Central Repository.

More recently, Benelallam \etal \cite{Benelallam2019} created the Maven Dependency Graph, an open source data set containing a snapshot of the whole Maven Central Repository taken on September 2018, stored in a temporal graph database modelling all dependencies.
This dataset has been used for various purposes, such as the study of dependency bloat~\cite{soto2021comprehensive} and diversity~\cite{Soto-Valero2019}.

\subsection{Mining Software Library Ecosystems}

Beyond the Java ecosystem, many software library ecosystems have been studied for a wide range of programming languages.
For the purpose of analysing the dependency networks of these ecosystems, \emph{Libraries.io} \cite{librariesio2020} has been used by several researchers (\eg \cite{zerouali2018empirical,decan:emse:2019,Stringer2020,zerouali2022impact,2022:icsr:makari}).
Five successive data dumps have been made available from 2017 to 2020, containing metadata from a wide range of different package managers.
No more recent data dumps have been released since Tidelift decided to discontinue active maintenance of the dataset.

As a kind of successor to \emph{Libraries.io}, the \emph{Ecosyste.ms} project\footnote{\url{https://ecosyste.ms}} was started in 2022.
Currently sponsored by the Open Collective\footnote{\url{https://opencollective.com}}, it focuses on expanding available data and APIs, as such providing a foundational basis for researchers to better analyze open source software, and for funders to better prioritize which projects need to be funded most.
The \emph{Ecosyste.ms} platform provides a shared collection of openly accessible services to support, sustain, and secure critical open source software components.
Each service comes with an openly accessible JSON API to facilitate the creation of new tools and services.
The APIs and data structures are designed to be as generic as possible, to facilitate analysing different data sources in an ecosystem-agnostic way.
Some of the supported services include:
\begin{itemize}
    \item An index of several millions of open source packages from dozens of package registries (for programming languages and Linux distributions), with tens of thousands new package versions being added on a daily basis.
    \item An index of the historical timeline of several billions of events that occurred across public git repositories (hosted on \github, \gitlab or \gitea) over many years, with hundreds of thousands of events being added on an hourly basis.
    \item An index of dozens of millions of open source repositories and \docker projects and their dependencies originating from a dozen of different sources, with tens of thousands new repositories being added on a daily basis.
    \item A range of services to provide software repository, contributor and security vulnerability metadata, parse software dependency and licensing metadata, resolve software package dependency trees, generate diffs between package releases, and many more.
\end{itemize}

\subsection{Mining Other Software Ecosystems}

Beyond the data sources mentioned above, a wide variety of other initiatives to mine, analyse or archive software ecosystems have been proposed through a plethora of datasets or data sources that are --or have been-- available for researchers or practitioners.

Of particular relevance is the \emph{Software Heritage} ecosystem~\cite{swhipres2017}.
It is the largest public software archive,
containing the development history of billions of source code files from more than 180 million collaborative software development projects.
Supported by a partnership with UNESCO, its long-term mission is to collect, preserve, and make easily accessible the source code of publicly available software.
It comes with its own filesystem~\cite{swh-fuse-icse2021} and graph dataset~\cite{msr-2020-challenge}.
%For more details we refer to \chap{SWH}, which is entirely focused on this ecosystem.

\emph{World of Code} (WoC) \cite{mockus2019woc, Ma2021WoC} is another ambitious initiative to create a very large and frequently updated collection of historical data in OSS ecosystems.
The provided infrastructure facilitates the analysis of technical dependencies, social networks, and their interrelationships.
To this end, WoC provides tools for efficiently correcting, augmenting, querying, and analysing that data ---a foundation for understanding the structure and evolution of the relationships that drive OSS activities.

\emph{Boa}~\cite{dyer2013boa,dyer2015boa,Boa2020MSR} is yet another initiative to support the efficient mining of large-scale datasets of software repository data.
Boa provides a domain-specific language and distributed computing infrastructure to facilitate this.

%\tom{@Coen: list of datasets to be further processed from here. I don't know if we need a separate bullet for each, and I am very likely to have missed some other relevant ones. On the other hand, we cannot list everyhing.}
Many other attempts have been made in the past to create and support publicly available software datasets and platforms, but these are no longer actively maintained today.
We mention some notable examples below.
The \emph{PROMISE} Software Engineering Repository is a collection of publicly available datasets to serve researchers in conducting predictive software engineering  in a  repeatable, verifiable and refutable way~\cite{PROMISE2005}. %\tom{(2004-2006)Probably not that interesting since no longer maintained since 2006...}
\emph{FLOSSmole} is another collaborative collection of OSS project data~\cite{FLOSSmole}. %\tom{No longer maintained since 2017 so probably not relevant to mention either?}
\emph{Candoia} is a platform and ecosystem for building and sharing software repository mining tools and applications~\cite{Candoia2016, 7962355}. %\tom{No longer maintained since 2017 so perhaps not relevant to mention either?}
\emph{Sourcerer} is a research project aimed at exploring open source projects, and provided an open source infrastructure and curated datasets for other researchers to use~\cite{Sourcerer2014}.

\emph{DebSources} is a dataset containing source code and metadata spanning two decades of history related to the Debian Linux distribution until 2016~\cite{debsources-ese-2016}.

Jira is one of the most popular issue tracking systems (ITSs) in practice.
A first Jira repository dataset was created in 2015, containing more than 700K issue reports and more than 2 million issue comments extracted from the Jira issue tracking system of the Apache Software Foundation, Spring, JBoss and CodeHaus OSS communities \cite{Ortu2015}.
A more recent dataset created in 2022 gathers data from 16 public Jira repositories containing 1822 projects and spanning 2.7 million issues with a combined total of 32 million changes, 9 million comments, and 1 million issue links \cite{Montgomery2022Jira,montgomery2022alternative}.

%%%%%%%%%%%%%%%%%%%%%%%
%%%%%%%%%%%%%%%%%%%%%%%
\section{The CHAOSS Project}
\label{INT:sec:chaoss}

In an introductory chapter on software ecosystems it is indispensable to also mention the CHAOSS initiative (which is an acronym for Community Health Analytics in Open Source Software)~\cite{Goggins2021CHAOSS}.\footnote{\url{https://chaoss.community}}
It is a Linux Foundation project aimed at better understanding OSS community health on a global scale~\cite{Goggins2021}.
Unhealthy OSS projects can have a negative impact on the community involved in them, as well as on organisations that are relying on them.
CHAOSS therefore focuses on understanding and supporting health through the creation of metrics, metrics models, and software development analytics tools for measuring and visualising community health in OSS projects.

Two main OSS tools are proposed by CHAOSS to do so: \emph{Augur} and \emph{GrimoireLab}~\cite{GrimoireLab2021}.
The latter is an open source toolkit with support for extracting, visualising, and analysing activity, community, and process data from 30+ data sources related to code management, issues, code reviewing, mailing list, developer fora and more.
Perhaps one shortcoming of these tools is that they have not been designed to scale up to visualise or analyse health issues at the level of ecosystems containing thousands or even millions of interconnected projects.

\section{Summary}

%\tom{@coen, here is a first stab towards a wrap-up / summary / conclusion for this chapter?}

This introductory chapter served as a first stepping stone for newcomers in the research field of software ecosystems.
We aimed to provide the necessary material to get up to speed in this domain.
After a historical account of where software ecosystems originated from, we highlighted the different perspectives on software ecosystems, and their accompanying definitions.
We categorised the different kinds of software ecosystems, providing many examples for each category.

%Since the book to which this introductory chapter belongs focuses on software ecosystem tooling and analytics, 
We presented a rich set of data sources and datasets that have been or can be used for mining software ecosystems.
Given that the field of software ecosystems is evolving at a rapid pace, it is difficult to predict the direction into which it is heading, and the extent to which the current tools and data sources will evolve or get replaced in the future.

%\bibliographystyle{spmpsci}
%One single bib file containing all references!
%\bibliography{references.bib}

\end{document}